\documentclass[aps,amsmath,showpacs,amsfonts,11pt]{revtex4}
\usepackage{graphicx}
\usepackage{epstopdf}
\usepackage{epsfig,graphicx}
\usepackage[english]{babel}
\usepackage{amsfonts}
\usepackage{amsmath}
\usepackage{latexsym}
\usepackage{graphics,bm}
\usepackage{dcolumn}
\usepackage{bm}
\usepackage{rotating}
\begin{document}

\title{Photon statistics of radiation emitted by two quantum wells embedded in two optically coupled semiconductor microcavities}

\author{ Shahnoor Ali$^{1}$ and Aranya B Bhattacherjee$^{2}$ }

\address{$^{1}$School of Physical Sciences, Jawaharlal Nehru University, New Delhi-110067, India} 
\address{$^{2}$Department of Physics, Birla Institute of Technology and Science, Pilani,
Hyderabad Campus,  Telangana State - 500078, India}

\begin{abstract}
We study theoretically the photon statistics of the field emitted from two optically coupled semiconductor microcavities each containing a quantum well. The emission is determined by the interplay between exciton-photon coupling in each quantum well and internal interaction between the two optically coupled microcavities. A high degree of coherent control and tunability via the quantum well-cavity coupling strength and photon tunneling over the photon statistics of the transmitted field can be achieved. We demonstrate that the optical property of radiation emitted by one quantum well can be controlled by the properties of the second quantum well. This result has the potential to be used in quantum information processing. We show that the exciton-photon coupling influences the polariton resonances in the intensity spectrum of the transmitted field. The results obtained in this investigation has the potential to be used for designing efficient controllable all-optical switch and high sensitive optical sensor.  

\smallskip
\noindent \textbf{Keywords.} Quantum well, Semiconductor micro-cavity, Photon statistics
\end{abstract}

\maketitle

\section{Introduction}

Optical properties of semiconductor nanostructures like quantum wells (QW) and quantum dots (QD) offer many new fascinating features \citep{shields, ghosh, khitrova, haicher, richard, vahala, deveaud, savona,gibbs} with potential applications in optoelectronic devices \citep{shields}. In this regard, the formation of an electron-hole pair termed as exciton plays a crucial role. The exchange of energy between the excitons and the vacuum field is attributed to the observed quantum optical response of QW and QD. The interaction of an exciton in a QW with optical modes of a micro-cavity has been studied extensively in the past \citep{sete1, sete2, sete3, sete4, sete5}. In semiconductor nanostructures embedded in micro-cavities, such coherent exchange of energy becomes observable as vacuum Rabi splitting in the strong coupling regime \citep{tassone,yamamoto,weisbuch,pau,jacobson,reithmaier,ishida,santhosh}. A strong coupling is achieved when the exciton-field coupling strength is much larger than the relaxation rates of the medium and of the cavity \citep{gibbs, hennessy}. The coherent energy exchange between excitons and photons can be explained as the formation of polaritons which are the mixed modes of QW exciton and cavity photon \citep{pau, chen, sermage}.
QW and QD embedded in photonic crystal cavities are considered highly attractive candidates for implementing optoelectronic devices such as an all optical switch \citep{majumdar, kiraz, faraon, lodahl, bhattacharya, bhattacherjee, heshami} which has been demonstrated in recent experiments \citep{dory, englund}. To make such optoelectronic device a reality, complete coherent control of the quantum device is essential. 
In light of these interesting quantum optical features associated with semiconductor nanostructures in microcavities and possible new optoelectronic applications, we investigate in the current paper a relevant question: How is the quantum optical property of the radiation emitted from a QW in a semiconductor micro-cavity affected in the presence of a second QW?  At the heart of quantum information processing is conditional quantum dynamics, where measurements made on one quantum system is controlled by the quantum state of another system. Such conditional dynamics in interacting quantum dots have been realized experimentally \citep{robledo}. In particular, we will investigate the radiation emitted from two micro-cavities each containing a QW. The two micro-cavities are coupled due to photon tunneling. We will investigate the steady state mean cavity photon number, the dynamical evolution of the intensity of fluorescent light and the intensity spectrum of the transmitted field. To this end, we will be using both analytical as well as numerical tools.

\section{System Hamiltonian and steady state}

\begin{figure}[ht]
\hspace{-1.3cm}
\includegraphics [scale=0.3]{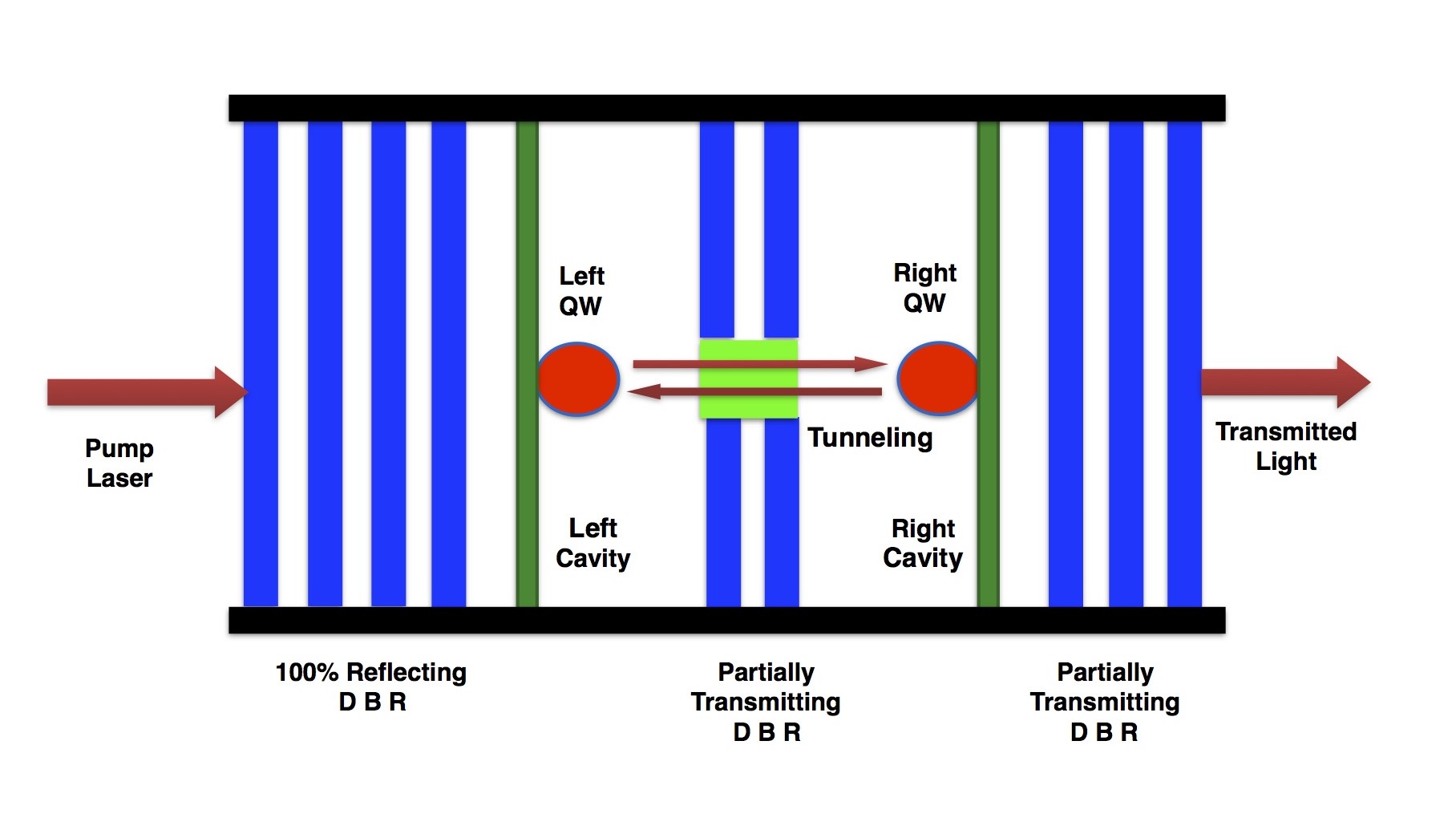}  
\caption{(Color online) Schematic representation of the setup studied in the text. It consists of two semiconductor micro-cavities made up of distributed Bragg reflectors (DBR) as shown. The left micro-cavity is driven by a strong pump laser. Both the cavities confine a quantum well each which is coupled to the respective field mode.  The two micro-cavities are optically coupled via the tunneling of photons between the two cavitiies. The blue and white strips correspond to AlGaAs and GaAs layers respectively.}
\label{f1}
\end{figure}

We consider a system consisting of two coupled micro-cavities, each containing a single semiconductor quantum well and supporting a field mode as shown in fig.1. Experimentally our proposed system could be InAs quantum well in GaAs semiconductor micro-cavity. These cavities are formed with the help of a set of distributed Bragg reflectors (DBR). In addition to this, photons are injected into the left cavity through an external pump.  Photons are able to tunnel between these two cavities. Thus the right cavity is driven by the output optical field from the left cavity. The field modes of the two micro-cavities thus constructed are coupled to the exciton mode of their respective QW, i.e the left micro-cavity mode is coupled to the left QW exciton mode while the right micro-cavity mode is coupled to the right QW exciton mode.

An exciton in the QW can be considered as a quasi-particle resulting from the interaction between one hole in the valence band and one electron in the conduction band. In the weak excitation regime, where the density of the excitons is sufficiently low, the interaction between the excitons due to coulomb interaction is extremely weak and thus can be ignored. We can treat the exciton as a composed boson when the exciton radius is significantly smaller than the average separation between neighbouring excitons. The left cavity is driven at rate $\lambda= \sqrt{\frac{2 P \kappa_{L}}{\hbar \omega_{p}}}$ through the left DBR by a laser with frequency $\omega_{p}$ and power $P$. The left cavity decay rate is $\kappa_{L}$. The pump is assumed to excite a single mode of the left cavity with frequency $\omega_{L}$. The coupled exciton-optical system is described by the Hamiltonian in a frame rotating with the pump frequency $\omega_{p}$ as,

\begin{eqnarray}
H &=&\Delta_{L} a_{L}^{\dagger}  a_{L}+\Delta_{R} a^{\dagger}_{R}a_{R}+\Delta \Omega_{1} c^{\dagger}_{L} c_{L}+\Delta \Omega_{2} c^{\dagger}_{R} c_{R}+J (a^{\dagger}_{L} a_{R}+a^{\dagger}_{R} a_{L}) \\ \nonumber
&+& i G_{1} (a^{\dagger}_{L} c_{L}-c^{\dagger}_{L} a_{L}) + i G_{2} (a^{\dagger}_{R} c_{R}-c^{\dagger}_{R} a_{R}) +i \lambda (a^{\dagger}_{L}-a_{L}).
\end{eqnarray}

\bigskip

Here $a_{L}$ and $a_{R}$ are the annihilation operators for a photon in the left and right micro-cavity respectively. The operators $c_{L}$ and $c_{R}$ are the annihilation operators for an exciton in the left and right QW respectively. Here $\Delta_{L}= \omega_{L}-\omega_{p}$, $\Delta_{R}=\omega_{R} - \omega_{p}$, $\Delta \omega_{1} = \omega_{1}-\omega_{p}$ and $\Delta \omega_{2}=\omega_{2}-\omega_{p}$. The left and right cavity frequencies are $\omega_{L}$ and $\omega_{R}$ respectively while $\omega_{1}$ and $\omega_{2}$ are the left and right exciton mode frequencies respectively. The fifth term in the Hamiltonian (Eq.1) describes the tunneling of the cavity photons between the two cavities with $J$ as the tunneling constant. The sixth and the seventh terms in the Hamiltonian describes the linear exciton-photon interactions with exciton-photon interaction strengths $G_{1}$ and $G_{2}$ for the left and right QW excitons  respectively. The last term describes the strong pump of amplitude $\lambda$.

Using the Hamiltonian (1) and taking into account the dissipation processes, one obtains the following quantum Langevin equations:

\begin{equation}
\frac{d a_{L}}{dt}=-(i \Delta_{L}+\kappa_{L})a_{L}-i J a_{R}+G_{1}c_{L}+\lambda+\sqrt{2 \kappa_{L}} a^{in}_{L},
\end{equation}

\begin{equation}
\frac{d a_{R}}{dt}=-(i \Delta_{R}+\kappa_{R})a_{R}-i J a_{L}+G_{2}c_{R}+\sqrt{2 \kappa_{R}} a^{in}_{R},
\end{equation}

\begin{equation}
\frac{dc_{L}}{dt}=-i \Delta \omega_{1} c_{L}-G_{1} a_{L}-\gamma_{L} c_{L}+ \sqrt{2 \gamma_{L}} c^{in}_{L},
\end{equation}

\begin{equation}
\frac{dc_{R}}{dt}=-i \Delta \omega_{2} c_{R}-G_{2} a_{R}-\gamma_{R} c_{R}+ \sqrt{2 \gamma_{R}} c^{in}_{R}.
\end{equation}

\bigskip

Here $\kappa_{L}$ ($\kappa_{R}$) is the left (right) cavity mode damping rate and $\gamma_{L}$ ($\gamma_{R}$) is the left (right) QW exciton spontaneous emission rate. Further $a^{in}_{L}$ ($a^{in}_{R}$) and $c^{in}_{L}$ ($c^{in}_{R}$) are the input vacuum noise whose correlation functions in the frequency domain are given by $\left< a^{in}_{i}(\omega) a^{\dagger in}_{i}(\omega')\right>= 2 \pi (1+n_{a i}) \delta(\omega-\omega') $, $\left< c^{in}_{i}(\omega) c^{\dagger in}_{i}(\omega')\right>= 2 \pi (1+n_{c i}) \delta(\omega-\omega') $, $\left< a^{\dagger in}_{i}(\omega) a^{ in}_{i}(\omega')\right>= 2 \pi n_{a i} \delta(\omega-\omega') $ and $\left< c^{\dagger in}_{i}(\omega) c^{ in}_{i}(\omega')\right>= 2 \pi n_{c i} \delta(\omega-\omega') $. Here $i=L,R$. Also $n_{ai}$ ($n_{ci}$) is the equilibrium photon (exciton) number in the $i^{th}$ cavity (QW).

\begin{figure}[ht]
\hspace{-1.3cm}
\begin{tabular}{cc}
\includegraphics [scale=0.60]{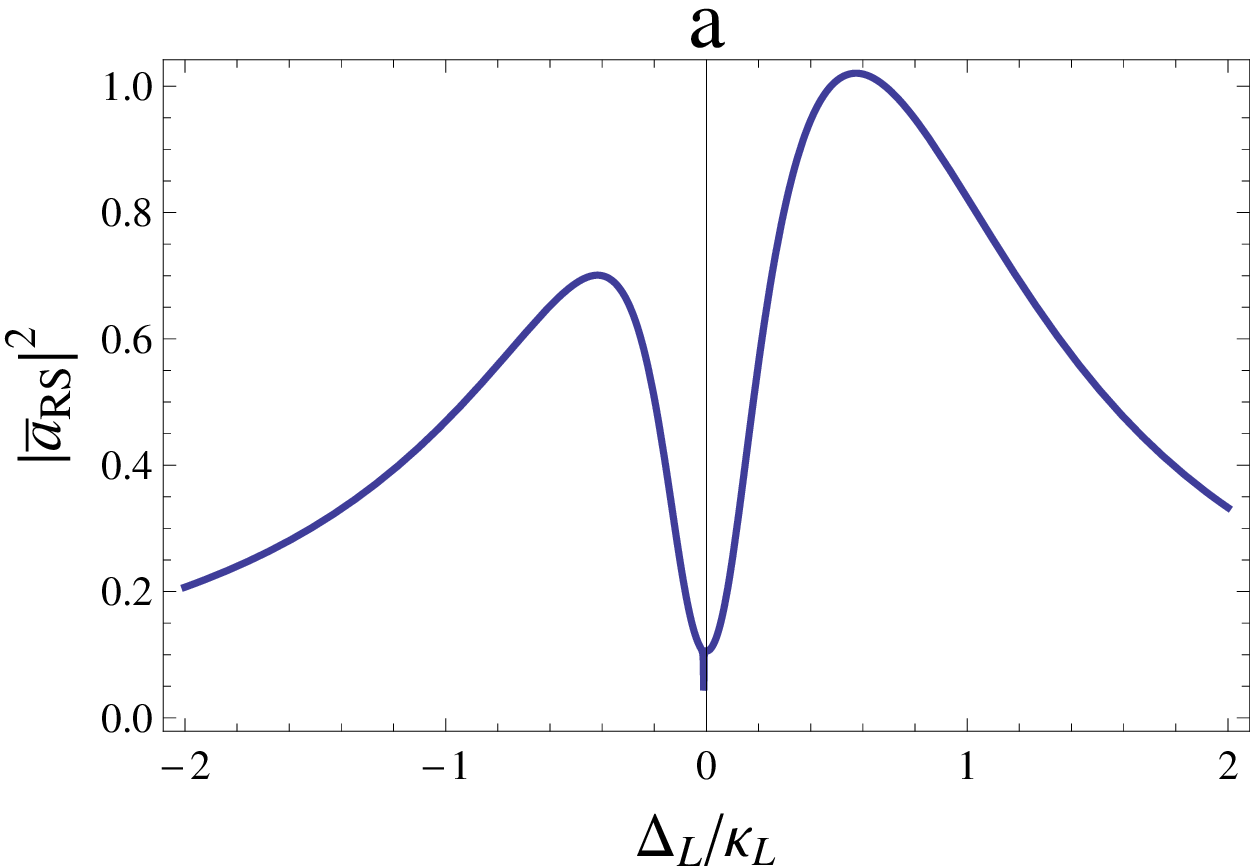}  \hspace{4mm} \includegraphics [scale=0.60] {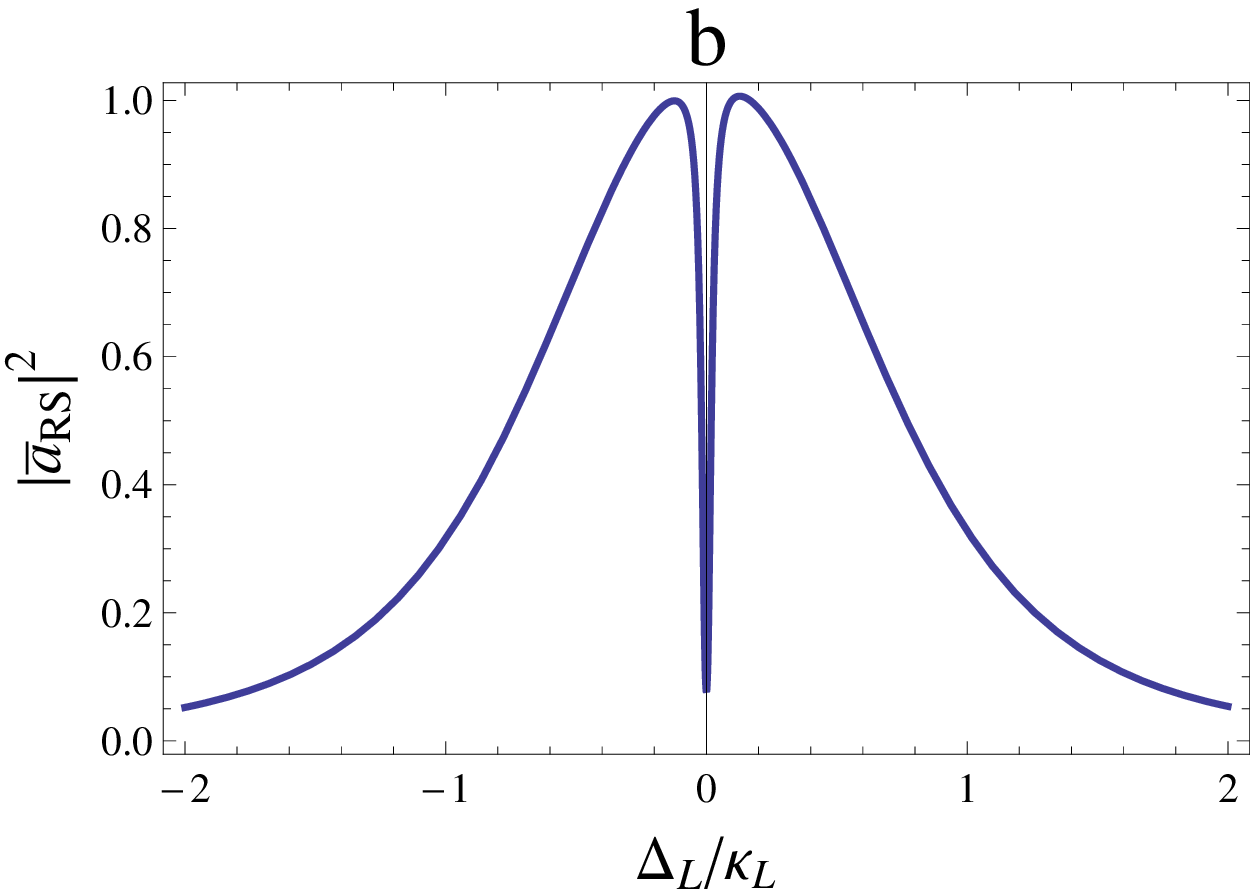}\\
 \end{tabular}
\caption{(Color online) Plots of the normalized mean intracavity photon number $|\bar{a}_{RS}|^{2}$ versus the dimensionless left cavity-pump laser detuning $\Delta_{L}/\kappa_{L}$ for the asymmetric case (a)
: $J=0.33$, $\kappa_{R}=0.8$, $\gamma_{L}=0.1$, $\gamma_{R}=0.2$, $\Delta_{R}=2.0$, $\Delta \omega_{1}=\Delta_{L}$, $\Delta \omega_{2}=-0.82$, $G_{1}=0.5$, $G_{2}=0.6$, $\lambda=10$ and symmetric case (b): $J=0.33$, $\kappa_{R}=0.8$, $\gamma_{L}=0.1$, $\gamma_{R}=0.2$, $\Delta_{R}=2.0$, $\Delta \omega_{1} = \Delta \omega_{2} = \Delta_{L}= \Delta_{R}$, $G_{1}=0.1$, $G_{2}=0.1$, $\lambda=4$. All frequencies are in the units of $\kappa_{L}$.} 
\label{f1}
\end{figure}

Using the Eqns.(2)-(5), we derive coupled equations for the macroscopic fields $\bar{a}_{L}$, $\bar{a}_{R}$, $\bar{c}_{L}$ and $\bar{c}_{R}$. These equations are obtained by replacing the operators with their corresponding classical values in the Heisenberg-Langevin equations (2-5). This replacement can be done in the high power limit $\lambda>1$.

\begin{equation}
\frac{d \bar{a}_{L}}{dt}=-(i \Delta_{L}+\kappa_{L}) \bar{a}_{L}-i J \bar{a}_{R}+G_{1} \bar{c}_{L}+\lambda,
\end{equation}

\begin{equation}
\frac{d \bar{a}_{R}}{dt}=-(i \Delta_{R}+\kappa_{R}) \bar{a}_{R}-i J \bar{a}_{L}+G_{2} \bar{c}_{R},
\end{equation}

\begin{equation}
\frac{d \bar{c}_{L}}{dt}=-i \Delta \omega_{1} \bar{c}_{L}-G_{1} \bar{a}_{L}-\gamma_{L} \bar{c}_{L},
\end{equation}

\begin{equation}
\frac{d \bar{c}_{R}}{dt}=-i \Delta \omega_{2} \bar{c}_{R}-G_{2} \bar{a}_{R}-\gamma_{R} \bar{c}_{R}.
\end{equation}

We now solve the Heisenberg equations (6-9) in the steady state and obtain the steady state solutions of $\bar{a}_{L}$ and $\bar{a}_{R}$ as,

\begin{equation}
\bar{a}_{RS}=\frac{-i J \lambda}{[(\kappa'_{R}+i \Delta'_{R})(\kappa'_{L}+i \Delta'_{L})+J^2]},
\end{equation}

\begin{equation}
\bar{a}_{LS}=\frac{ \lambda (\kappa'_{R}+i \Delta'_{R}) }{[(\kappa'_{R}+i \Delta'_{R})(\kappa'_{L}+i \Delta'_{L})+J^2]},
\end{equation}

where

\begin{equation}
\kappa'_{R}=\kappa_{R}+\frac{G_{2}^{2} \gamma_{R}}{(\gamma_{R}^{2}+\Delta \omega_{2}^{2})}
\end{equation}

\begin{equation}
\kappa'_{L}=\kappa_{L}+\frac{G_{1}^{2} \gamma_{L}}{(\gamma_{L}^{2}+\Delta \omega_{1}^{2})}
\end{equation}

\begin{equation}
\Delta'_{L}=\Delta_{L}-\frac{G_{1}^{2} \Delta \omega_{1}}{(\gamma_{L}^{2}+\Delta \omega_{1}^{2})}
\end{equation}

\begin{equation}
\Delta'_{R}=\Delta_{R}-\frac{G_{2}^{2} \Delta \omega_{2}}{(\gamma_{R}^{2}+\Delta \omega_{2}^{2})}
\end{equation}

\bigskip

We are now intrested in the variation of the steady state mean photon number $|\bar{a}_{RS}|^{2}$ in the right micro-cavity as a function of $\Delta_{L}$. The expression for $|\bar{a}_{RS}|^{2}$ is written as,

\begin{equation}
|\bar{a}_{RS}|^{2} = \frac{J^2 \lambda^{2}}{[(J^2+\kappa'_{R} \kappa'_{L}-\Delta'_{R} \Delta'_{L})^2+(\Delta'_{R} \kappa'_{L}+\Delta'_{L} \kappa'_{R})^2]}
\end{equation}

\bigskip

In Fig.2, we show the plot of $|\bar{a}_{RS}|^{2}$ as a function of $\Delta_{L}/\kappa_{L}$. The values of the parameters are based on earlier experimental studies \citep{shields, ghosh, khitrova, haicher, richard, vahala, deveaud, savona,gibbs,tassone,yamamoto,weisbuch,pau,jacobson,reithmaier,ishida,santhosh}. Clearly, plot of fig.2(a) is highly asymmetric around $\Delta_{L}=0$ and it also exhibits the switching characteristics of an all optical switch around $\Delta_{L}/\kappa_{L}=0$. The asymmetric structure of the split resonance is due to the asymmetric parameters chosen for the two QWs. If we chose the two micro-cavities and the two QWs to be identical then the split resonance structure exibits symmetrical behaviour around $\Delta_{L}/\kappa_{L}=0$ as shown in fig.2(b). It is also observed that the symmetric optical switching is very sharp compared to the asymmetric plot. This essentially shows that in order to design a very sensitive all optical switch, the two optical cavities and the two QWs should be identical.

\section{Photon Statistics}

In this section, we analyze the photon statistics of the fluorescent light of the right QW by calculating the intensity $ \left< c^{\dagger}_{R}(t) c_{R}(t) \right>$. In order to study the photon statistics, we adopt the mean-field as well as the full quantum model.

\begin{figure}[ht]
\hspace{-1.3cm}
\begin{tabular}{cc}
\includegraphics [scale=0.60]{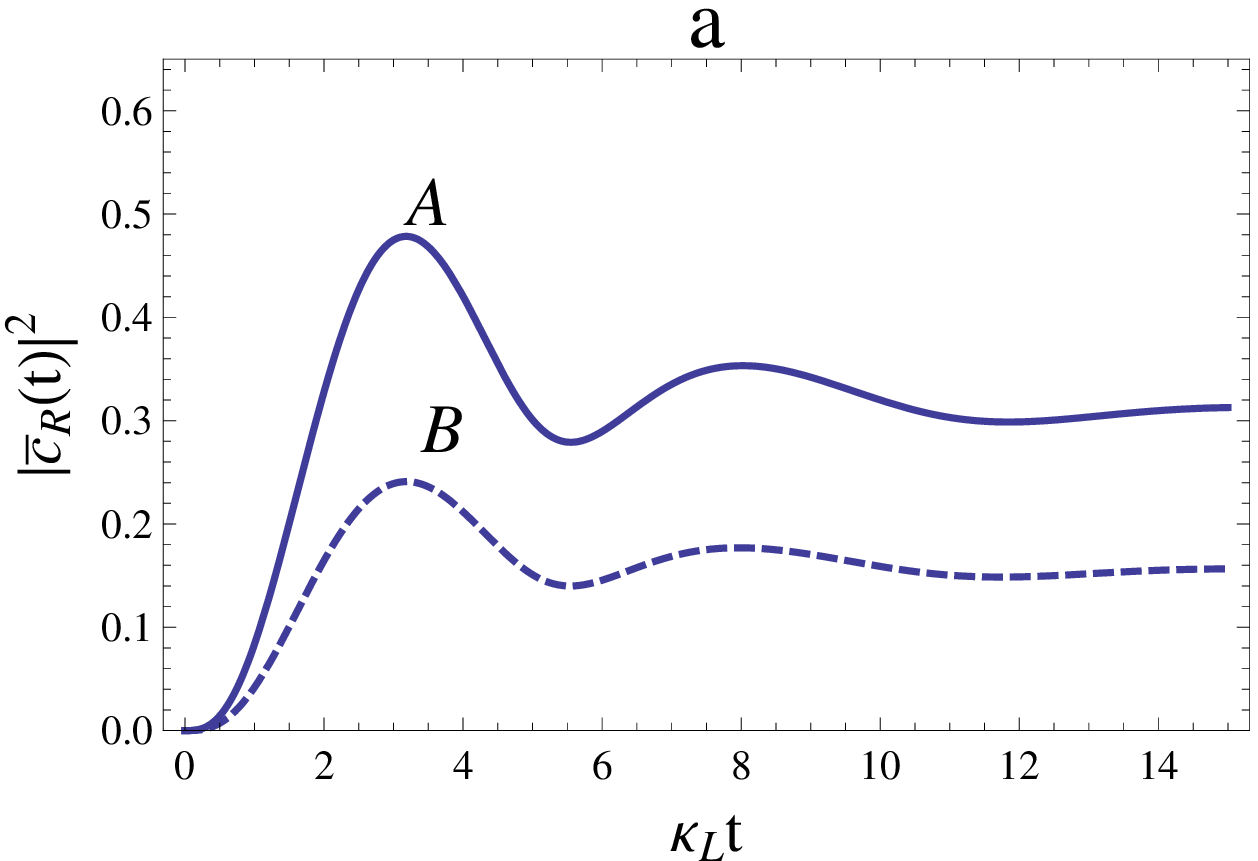}  \hspace{4mm} \includegraphics [scale=0.60] {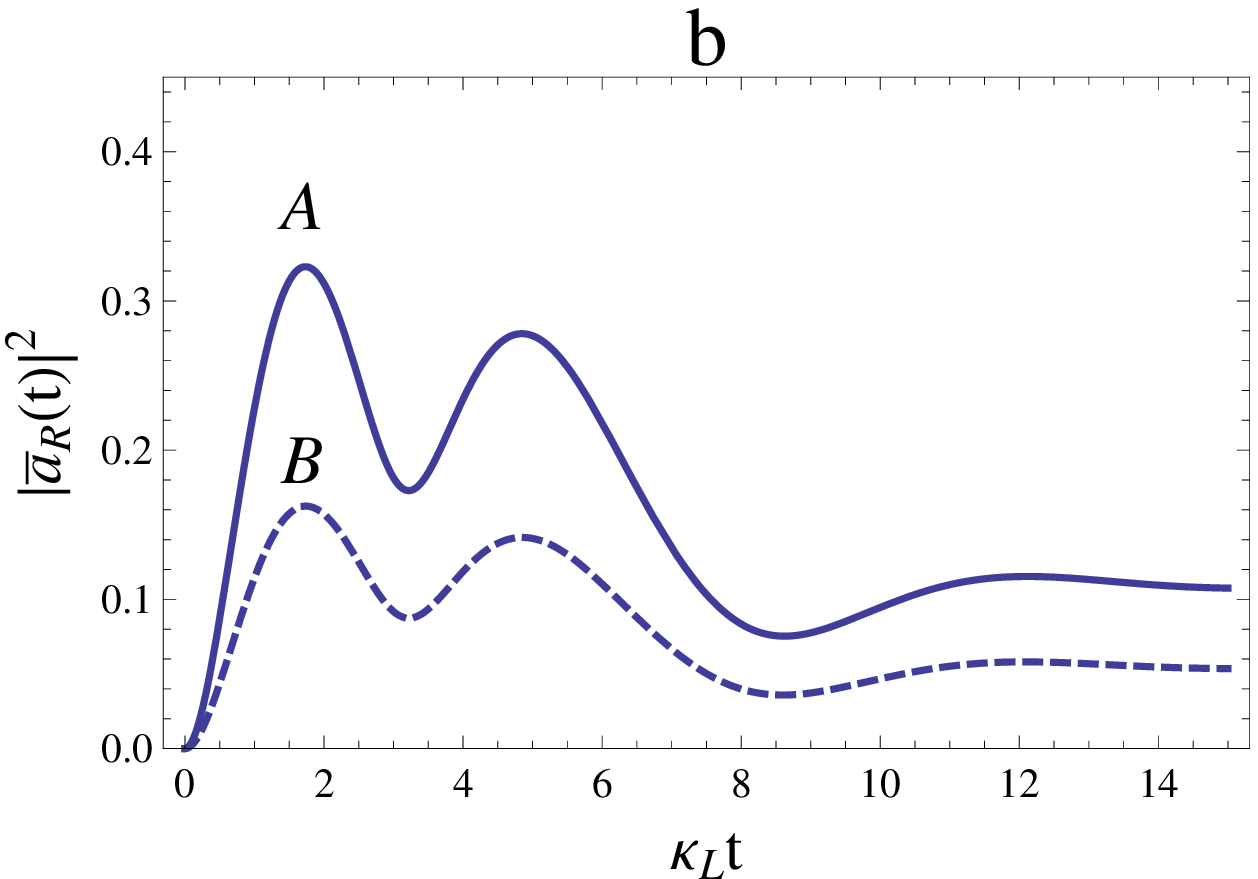}\\
 \end{tabular}
\caption{(Color online) Plot of the time dynamics of $|\bar{c}_{R}(t)|^{2}$ (plot (a)) and  right cavity photon number $|\bar{a}_{R}(t)|^{2}$ (plot (b)) as a function of $\kappa_{L} t$ for $\kappa_{R}=1.0$, $\gamma_{L}=0.1$, $\gamma_{R}= \gamma_{L}=0.1$, $\Delta_{R}=\Delta_{L}=0$, $\Delta \omega_{1}= \Delta \omega_{2}=0.3$, $G_{1}=G_{2}=0.9$, $\lambda=5$ for two different values of photon tunneling $J=0.2$ (A) and $J=0.1$(B).}
\label{f2}
\end{figure}

\begin{figure}[ht]
\hspace{-1.3cm}
\begin{tabular}{cc}
\includegraphics [scale=0.60]{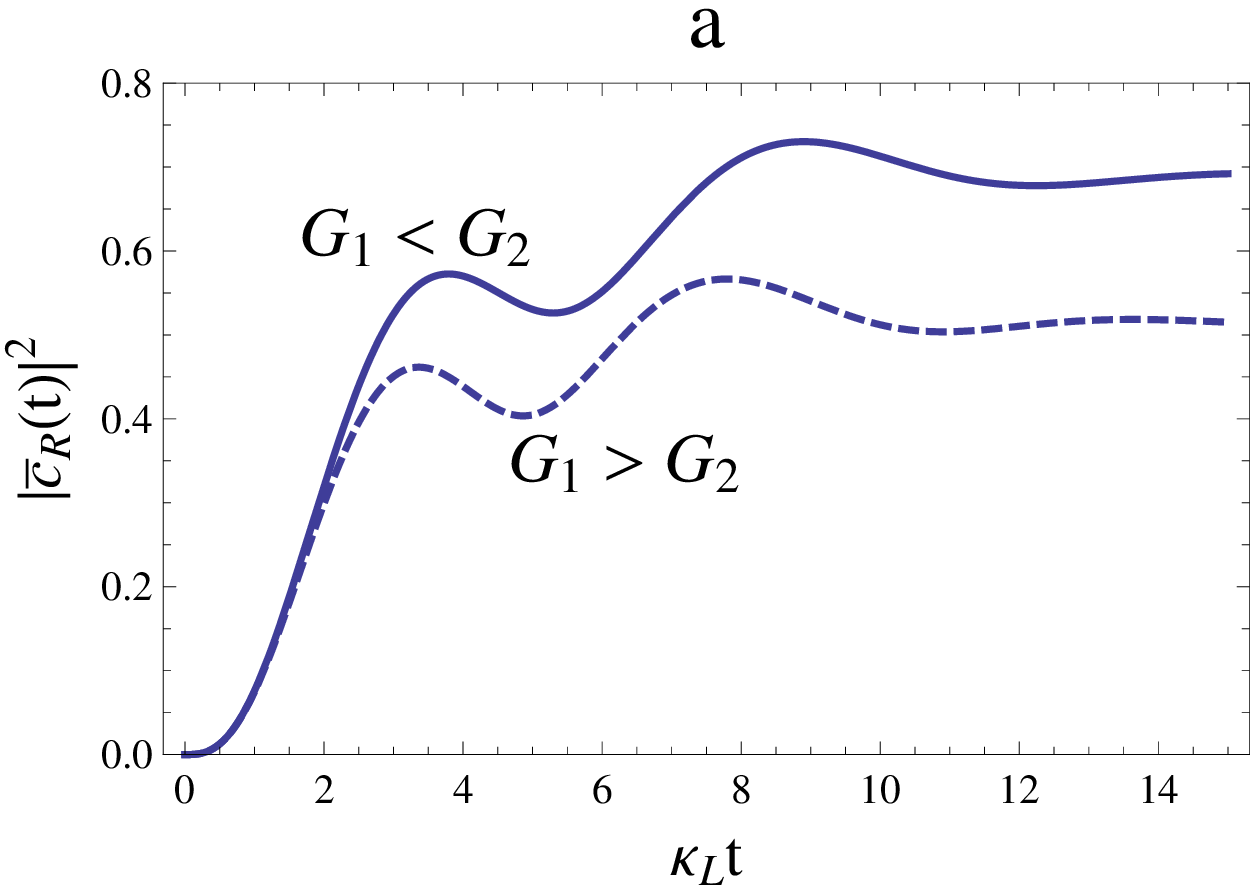} \hspace{4mm} \includegraphics [scale=0.60] {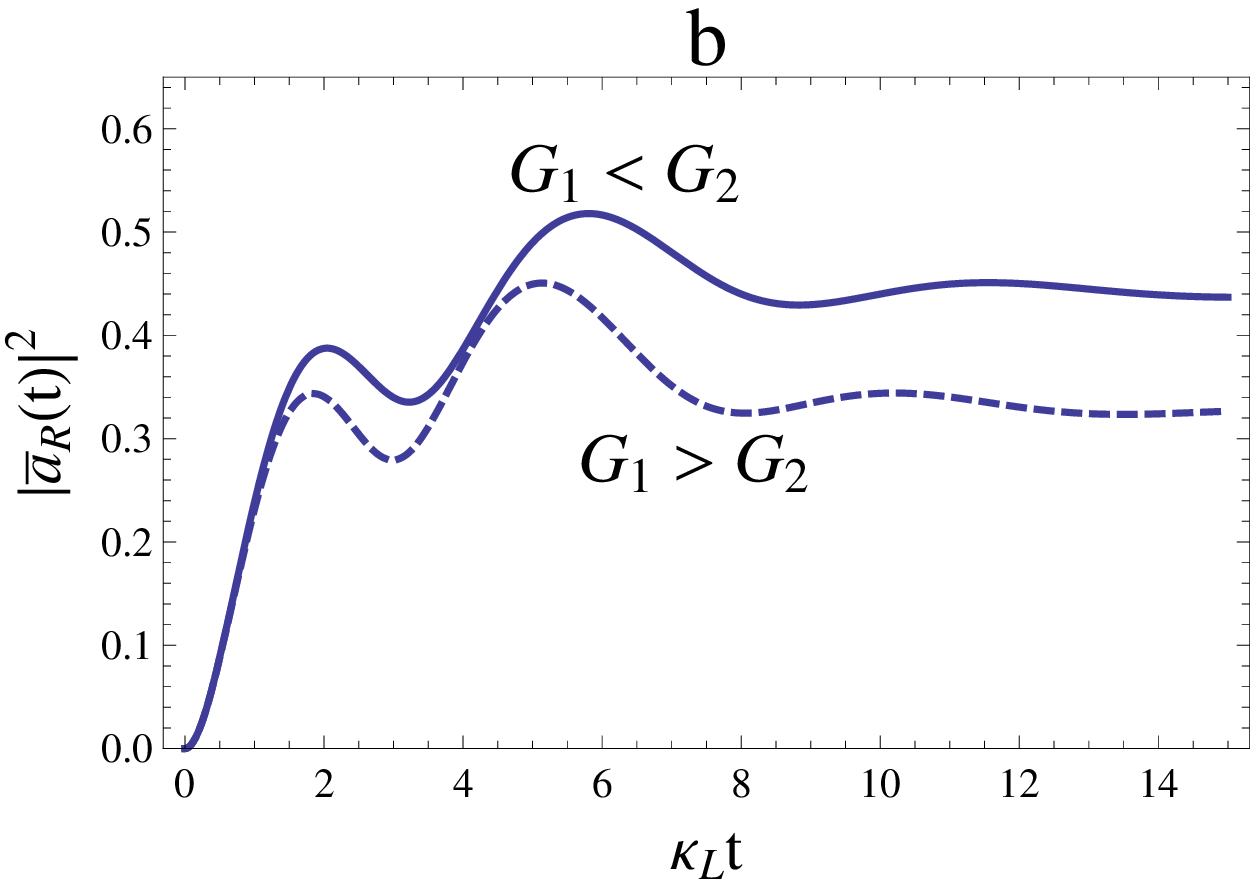}\\
 \end{tabular}
\caption{(Color online) Plot of the time dynamics of $|\bar{c}_{R}(t)|^{2}$ (plot (a)) and  right cavity photon number $|\bar{a}_{R}(t)|^{2}$ (plot (b)) as a function of $\kappa_{L} t$ for $\kappa_{R}=1.0$, $\gamma_{L}=0.1$, $\gamma_{R}= \gamma_{L}=0.1$, $\Delta_{R}=\Delta_{L}=0.2$, $\Delta \omega_{1}= \Delta \omega_{2}=0.5$, $J+0.2$,  $\lambda=5$ for two different set of exciton-photon couplings strengths $G_{1}$ and $G_{2}$. Solid line: $G_{1}=0.7, G_{2}=0.8$, Dashed line: $G_{1}=0.9, G_{2}=0.8$.} 
\label{f3}
\end{figure}

\subsection{Mean-Field Model}

In this approach, the Heisenberg equations of the motion for the classical cavity fields $\bar{a}_{L}$, $\bar{a}_{R}$ and excitons $\bar{c}_{L}$ , $\bar{c}_{R}$ (Eqns. 6-9) are solved numerically. The time dynamics of $|\bar{c}_{R}(t)|^{2}$ as a function of $\kappa_{L} t$ is shown in Fig.3(a) for two values of the tunneling parameter $J$ considering identical QWs and micro-cavities. The stronger the tunneling, larger is the amplitude of oscillations of $|\bar{c}_{R}(t)|^{2}$, indicating the presence of larger number of photons in the right cavity and hence an increase in the probability for the light in the right cavity to excite more excitons in the QW. However, the time it takes to reach the steady state value  $|\bar{c}_{R}(t \rightarrow \infty)|^{2}$ is same for both values of $J$. The corresponding time dynamics of the right cavity photon number $|\bar{a}_{R}(t)|^{2}$ is shown in Fig.3(b). A larger number of photons is evident for the larger of the two values of $J$.
In Fig.4(a), we show the dynamics of $|\bar{c}_{R}(t)|^{2}$ for two values of $G_{1}$ with $G_{2}$ fixed. An interesting observation is that the amplitude of oscillations is higher when $G_{1}<G_{2}$. This can be explained by noting that under strong exciton-photon coupling, there is a high photon density (this is evident from the plot of $|\bar{a}_{R}(t)|^{2}$ in Fig.4(b)) and hence high number of excitons. Thus when $G_{1}<G_{2}$, the photon density in the right cavity will be more compared to the left cavity and hence the observed result.

\subsection{The Full Quantum Model: Master Equation Approach}

\begin{figure}[ht]
\hspace{-0.7cm}
\begin{tabular}{cc}
\includegraphics [scale=0.55]{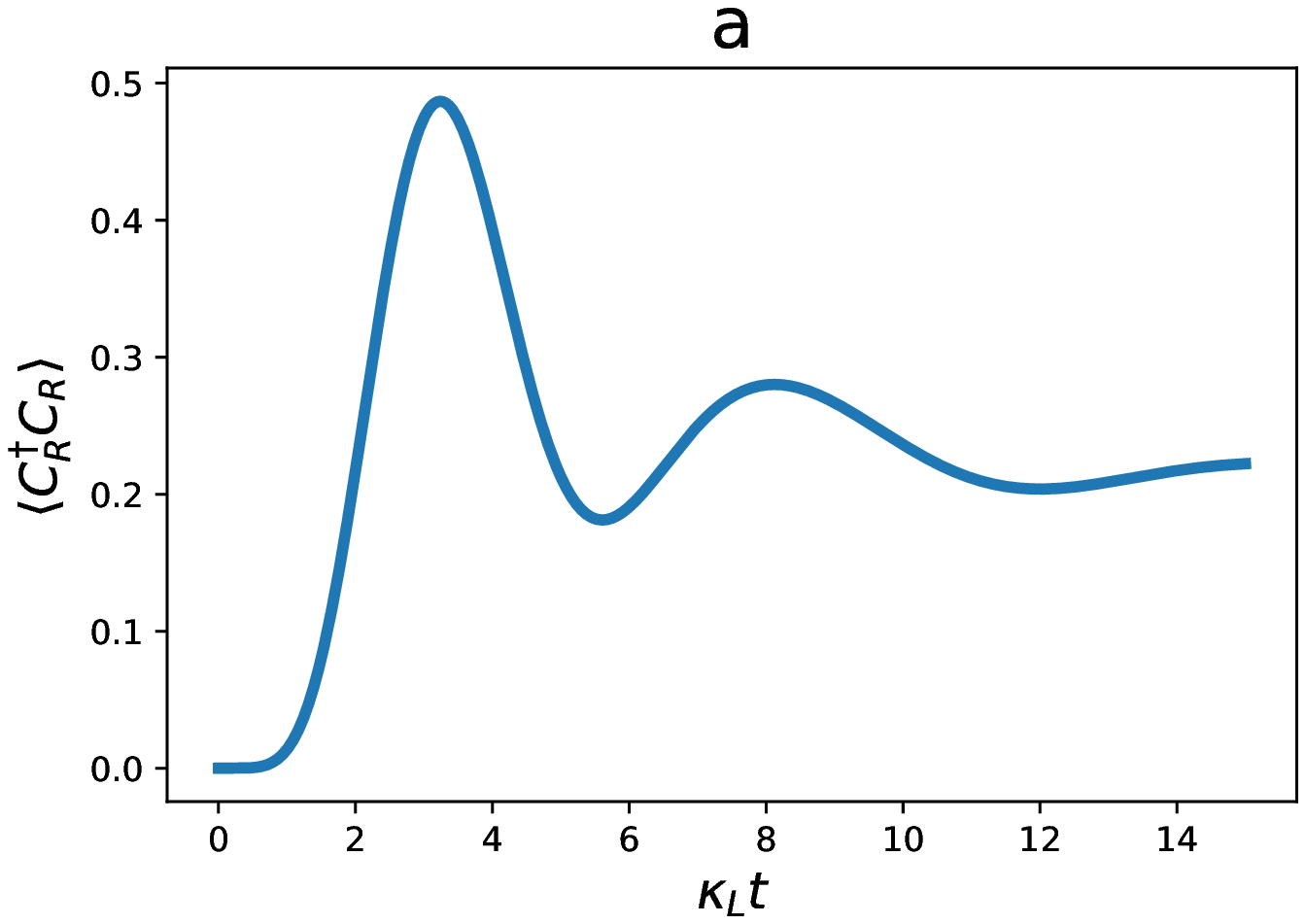} \includegraphics [scale=0.55] {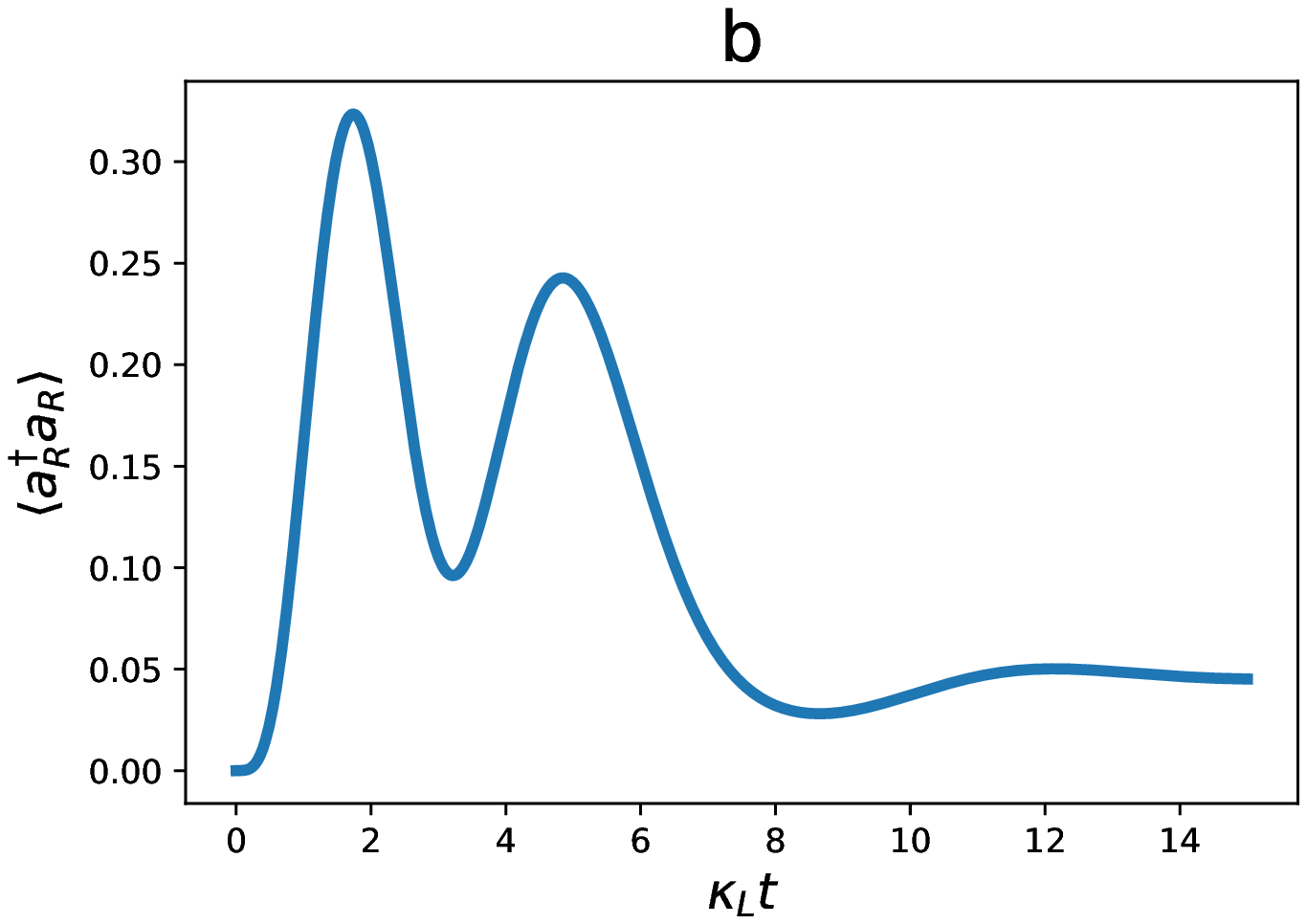}\\
 \end{tabular}
\caption{(Color online) Plot of the time dynamics of $\left< c^{\dagger}_{R}(t) c_{R}(t) \right>$ (plot (a)) and  right cavity photon number $\left< a^{\dagger}_{R}(t) a_{R}(t) \right>$ (plot (b)) as a function of $\kappa_{L} t$ in the frame of  master equation approach using the Quantum Toolbox in Python (QuTip, Release 4.2.0) corresponding to the parameters of Fig.(3), plot A.  } 
\label{f4}
\end{figure}

The intensity can be calculated numerically in the frame of master equations in the Lindblad form \citep{walls} using the Quantum Toolbox in Python (QuTip, Release 4.2.0) \citep{qutip}. Considering the dissipation of the cavity with decay rates $\kappa_{L}$, $\kappa_{R}$ and exciton exciton decay rates $\gamma_{L}$ and $\gamma_{R}$, the master equation of the dynamics of the system satisfies,

\begin{equation}
\frac{d\rho}{dt}=-i [H, \rho]+\kappa_{L}D[a_{L}]\rho+\kappa_{R} D[a_{R}] \rho+\gamma_{L} D[c_{L}] \rho+ \gamma_{R} D[c_{R}] \rho,
\end{equation}

where $\rho$ is the density matrix of the system and $H$ is the Hamiltonian (1). Here $D[O] \rho= \frac{1}{2}(2O\rho O^{\dagger}-O^{\dagger} O \rho-\rho O^{\dagger} O)$ is the Lindblad type of dissipation corresponding to the collapse operator $O$. The full quantum mechanical solution using the master equation approach in QuTip for $\left< c^{\dagger}_{R}(t) c_{R}(t)\right>$ and $\left< a^{\dagger}_{R}(t) a_{R}(t)\right>$ is shown in Fig5(a) and 5(b) respectively for $J=0.2 \kappa_{L}$. A comparison of Fig.5 with Fig.3 shows that the mean-field result is in excellent agreement with the full quantum model.

\section{Intensity spectrum of transmitted field}

In order to study the intensity spectra of the transmitted field, we first linearize the quantum Langevin equations (2-5) by rewritting the photon and exciton operators as the sum of their mean-field steady state values and the corresponding fluctuation operators. Thus we get $a_{L}=\bar{a}_{LS}+\delta a_{L}$,  $a_{R}=\bar{a}_{RS}+\delta a_{R}$,  $c_{L}=\bar{c}_{LS}+\delta c_{L}$ and $c_{R}=\bar{c}_{RS}+\delta c_{R}$. Here $\delta a_{L}$, $\delta a_{R}$, $\delta c_{L}$ and $\delta c_{R}$ are the fluctuation operators.

After making the above substitutions, the linearized Langevin equations for the fluctuation operators are,

\begin{equation}
\frac{d \delta a_{L}}{dt} = -(i \Delta_{L}+\kappa_{L}) \delta a_{L}-i J \delta a_{R}+G_{1} \delta c_{L}+\sqrt{2 \kappa_{L}} a^{in}_{L},
\end{equation}

\begin{equation}
\frac{d \delta a_{R}}{dt} = -(i \Delta_{R}+\kappa_{R}) \delta a_{R}-i J \delta a_{L}+G_{2} \delta c_{R}+\sqrt{2 \kappa_{R}} a^{in}_{R},
\end{equation}

\begin{equation}
\frac{d \delta c_{L}}{dt}=-(i \Delta \omega_{1}+\gamma_{L} ) \delta c_{L}-G_{1} \delta a_{L}+ \sqrt{2 \gamma_{L}} c^{in}_{L},
\end{equation}

\begin{equation}
\frac{d \delta c_{R}}{dt}=-(i \Delta \omega_{2}+\gamma_{R}) \delta c_{R}-G_{2} \delta a_{R}+ \sqrt{2 \gamma_{R}} c^{in}_{R},
\end{equation}

\bigskip

In order to calculate the intensity spectra of the transmitted field, we work in the frequency domain. We are interested in the intensity spectra of the field transmitted from the right side cavity. To this end the intensity spectrum of the field transmitted from the right cavity is written as

\begin{equation}
S_{R}(\omega) = 2 \kappa_{R} \int_{-\infty}^{\infty} \left< \delta a^{\dagger}_{R} (t+\tau) \delta a_{R}(t) \right> e^{-i(\omega-\omega_{o}) \tau} d \tau = 2 \kappa_{R} C_{R}(\omega).
\end{equation}

\bigskip

Here $2 \pi C_{R}(\omega) \delta (\omega+\omega')= \left< \delta a^{\dagger}_{R}(\omega) \delta a_{R} (\omega) \right>$. Making use of the correlation properties for the noise forces, the intensity spectrum $S_{R}(\omega)$ is written assuming $\kappa_{L}=\kappa_{R}=\kappa$ and $\gamma_{L}=\gamma_{R}=\gamma$ in the form,

\begin{equation}
S(\omega)=\frac{2 \kappa ( f_{1}(\omega)+f_{2}(\omega)+f_{3}(\omega)+f_{4}(\omega)  )}{[(\phi_{1r} \phi_{2r}-\phi_{1i}\phi_{2i}+J^2 \psi_{r})^2+(\phi_{1r} \phi_{2i}+\phi_{1i}\phi_{2r}+J^2 \psi_{i})^2]},
\end{equation}

\bigskip

where the expressions for $f_{1}(\omega)$, $f_{2}(\omega)$, $f_{3}(\omega)$, $f_{4}(\omega)$, $\phi_{1r}$, $\phi_{2r}$, $\phi_{1i}$, $\phi_{2i}$, $\psi_{r}$ and $\psi_{i}$ are explicitly given in Appendix A. 

We will explore the intensity spectrum of the transmitted field in strong, intermediate, weak and extremely weak coupling regimes. In order to understand the transmitted field we write the evolution matrix corresponding to the fluctuation equations (18-21),

\begin{figure}[ht]
\hspace{-1.3cm}
\begin{tabular}{cc}
\includegraphics [scale=0.60]{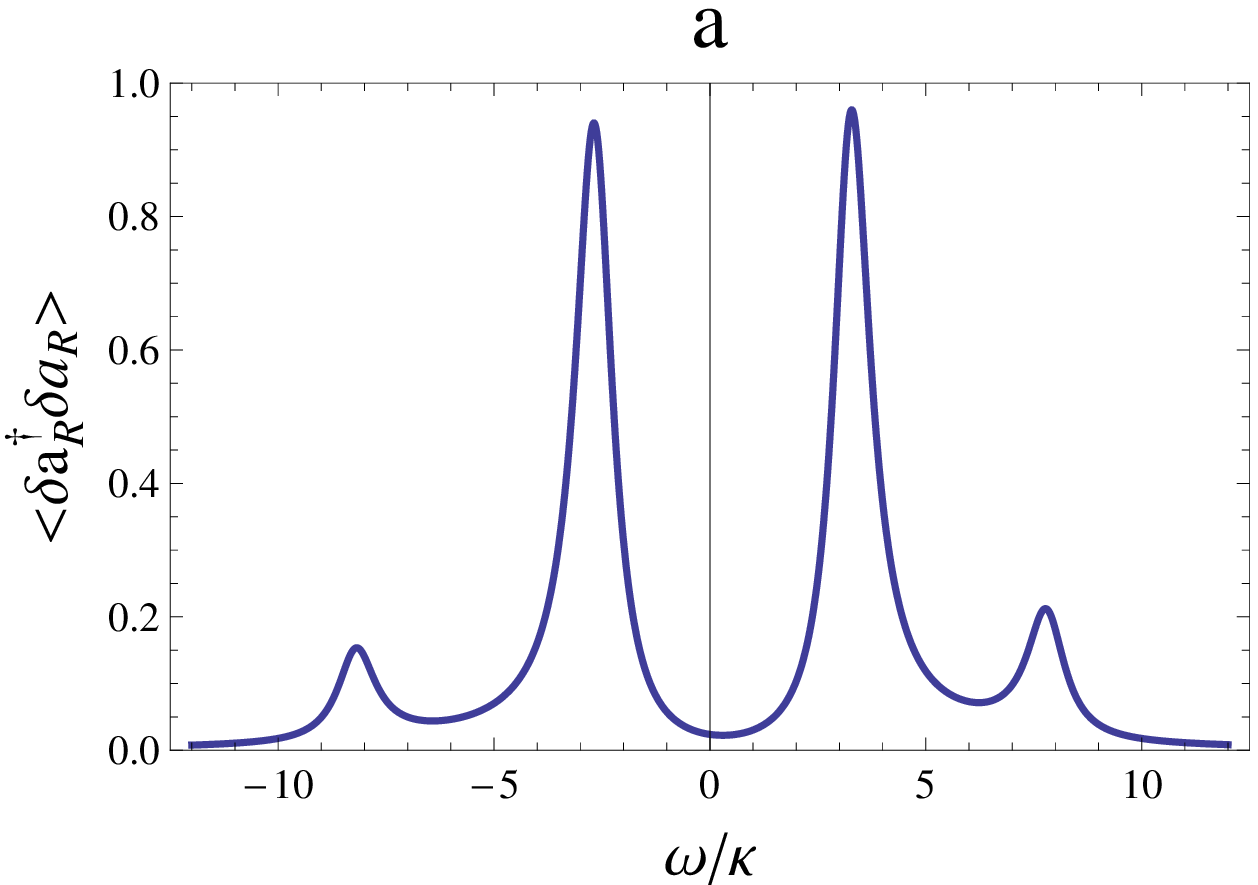} \hspace{4mm} \includegraphics [scale=0.60] {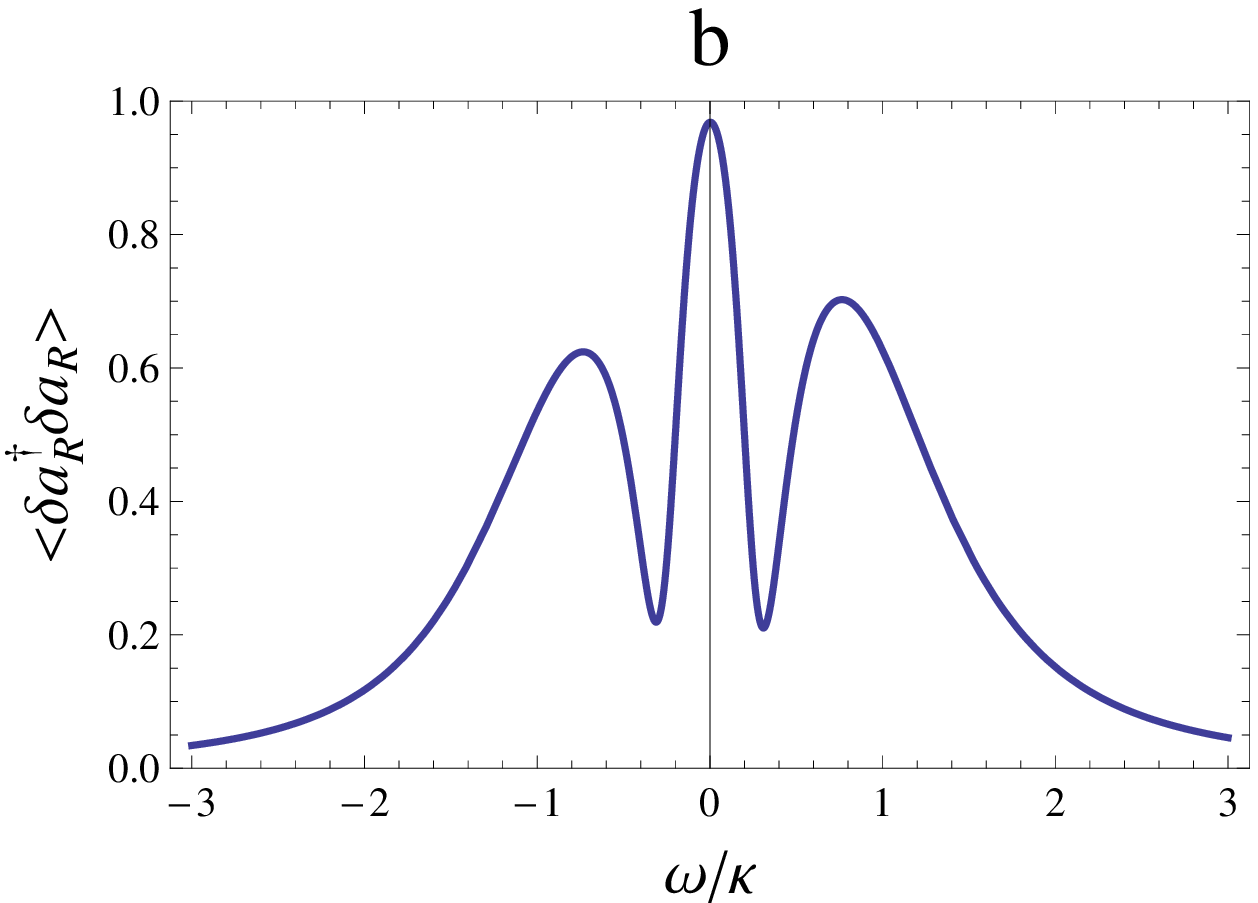}\\
\includegraphics [scale=0.61]{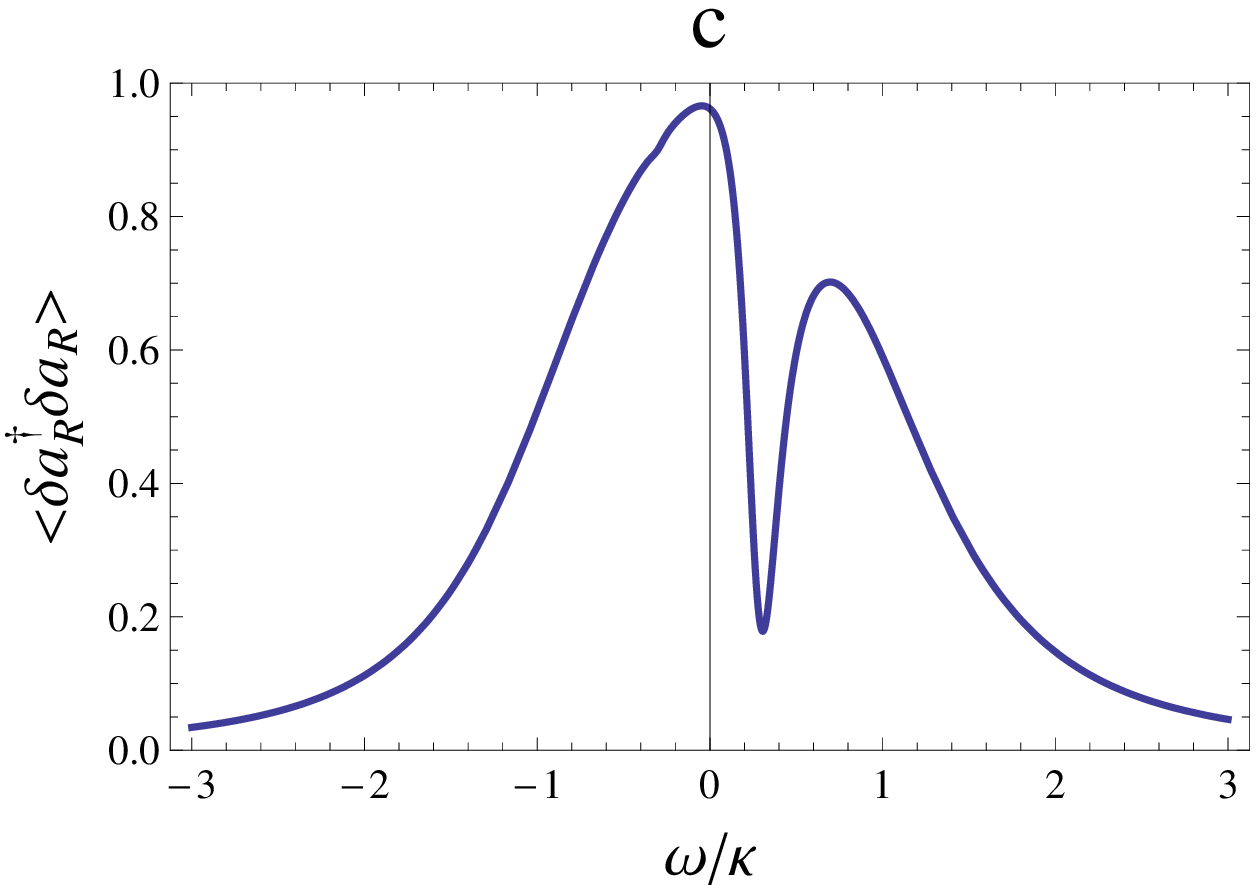} \hspace{4mm} \includegraphics [scale=0.60] {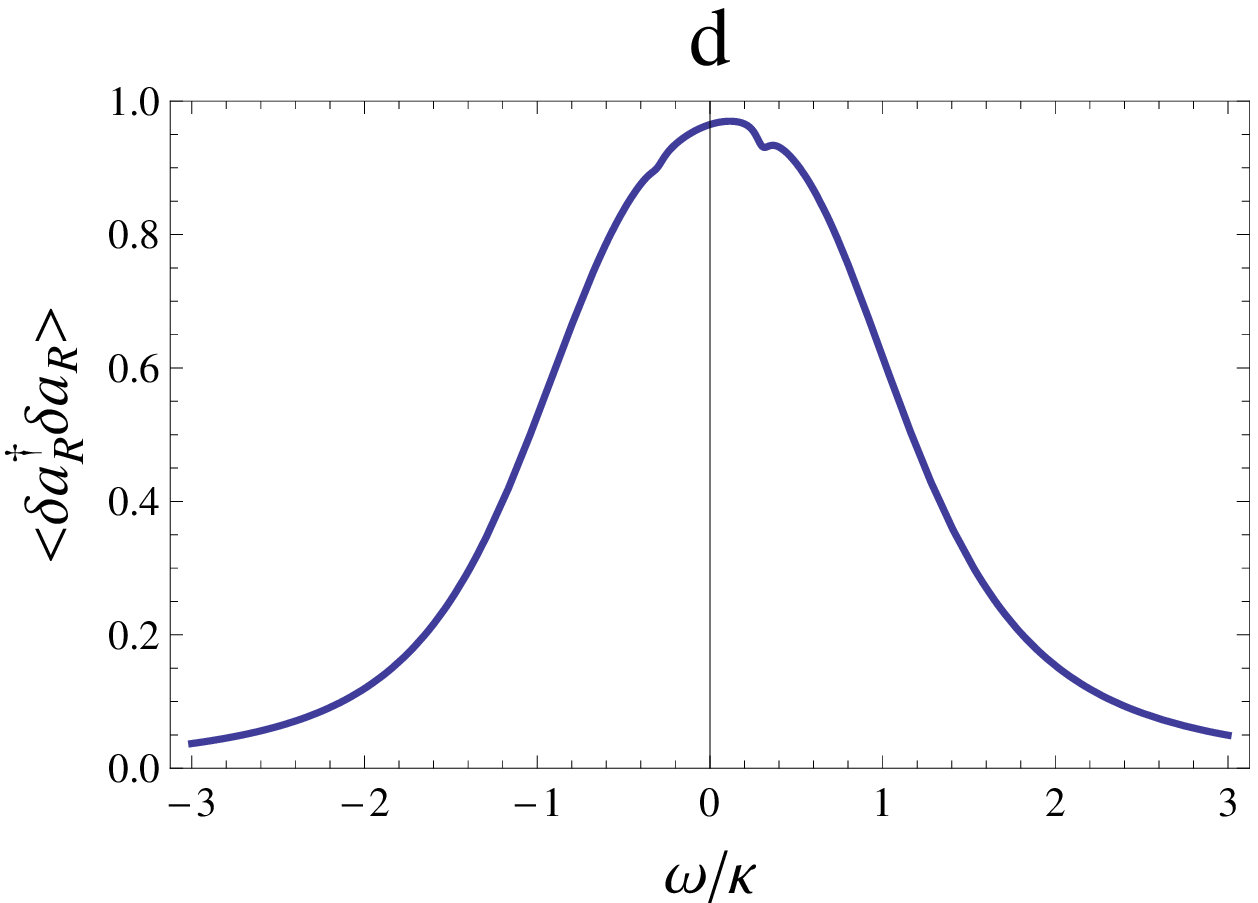}\\
\end{tabular}
\caption{(Color online) Plots of the normalized intensity spectra of the field transmitted from the right side cavity $S_{R}(\omega) $ for (a)  $\Delta \omega_{1}=-0.2, \Delta \omega_{2}=0.3, \Delta_{L}=-0.2, \Delta_{R}=0.3, G_{1}=6.5, G_{2}=3.3$, (b)   $\Delta \omega_{1}=-0.3$, $\Delta \omega_{2}=0.3$, $\Delta_{L}=-0.5$, $\Delta_{R}=0.5$, $G_{1}=0.4$, $G_{2}=0.4$, (c) $\Delta \omega_{1}=-0.2$, $\Delta \omega_{2}=0.3$, $\Delta_{L}=-0.2$, $\Delta_{R}=0.3$, $G_{1}=0.02$, $G_{2}=0.3$, (d) $\Delta \omega_{1}=-0.2$, $\Delta \omega_{2}=0.3$, $\Delta_{L}=-0.2$, $\Delta_{R}=0.3$, $G_{1}=0.02$, $G_{2}=0.03$. The equilibrium photon and exciton numbers $n_{ai}, n_{ci}<1$ $(i=L,R)$, $\gamma=0.05$ and $J=0.53$.}
\label{f5}
\end{figure}

\begin{equation}
M = \left(\begin{matrix}
      -(i \Delta_{L}+\kappa_{L}) & -i J & G_{1} & 0\\
      -i J &  -(i \Delta_{R}+\kappa_{R}) & 0 & G_{2}\\
      -G_{1} & 0 & -(i \Delta \omega_{1}+\gamma_{L}) & 0\\ 
       0 & -G_{2} & 0 &  -(i \Delta \omega_{1}+\gamma_{L})
     \end{matrix}\right).
\end{equation}

\bigskip

The transmitted field spectra should consist of four peaks corresponding to the four distinct eigenvalues of the evolution matrix $M$. Let us now study first the strong coupling regime ($G_{1}, G_{2}, J$ $>>$ $\kappa$, $\gamma$ and $G_{1}, G_{2} $ $>$ $J$). In the case of a single QW interacting strongly with a cavity mode, a strong exchange of photons between the cavity mode and the excitons takes place and hence the exciton-cavity coupled system (polariton) emission spectrum consists of two symmetric peaks \citep{sete2}. In our system there are two such subsystems (each subsystem is composed of one QW interacting with one cavity mode) and hence the transmitted spectrum should consist of four peaks. This is evident from the plot depicted in Fig.6(a)  in the strong coupling regime.
Let us now move into the intermediate coupling regime, $G_{1}, G_{2}, J$ $<$ $\kappa$ and $G_{1}, G_{2}, J$ $>$ $\gamma$. In the strong coupling regime, two distinct polariton resonances each corresponding to the two subsystems appeared. On the other hand in the intermediate coupling regime, the two independent polariton resonances mix and gives rise to the three peak structure in the transmitted spectra as shown in Fig. 6(b). The mixing of the two resonances (hybrid resonance) takes place since $J>\gamma$ and $J>G_{1},G_{2}$. The tunneling of the photons allows one subsystem to control the dynamics of the second subsystem. In the weak coupling regime, it is interesting to note that if any one of the exciton-photon coupling strengths $G_{1}$ or $G_{2}$ is less than $\kappa$ and $\gamma$, only two peaks appear. In such a parameter regime, one of the two polariton resonances disappear and the remaining two peaks corresponds to a hybrid resonance with major contribution coming from the subsystem for which the exciton-photon coupling is still larger than $\gamma$. The weak coupling case is depicted in Fig. 6(c). In the extremely weak coupling regime both $G_{1}$ and $G_{2}$ are less than $\kappa$ and $\gamma$. In such a situation, both the polariton resonances disappear and the emission spectrum displays only a single hydrid resonance as shown in Fig. 6(d).

Experimentally, this model can be realized as follows. Light confinement is achieved by the combined action of distributed Bragg reflectors (DBR) along the x-direction and air guiding dielectric provides confinement in the y-z plane \citep{m1}. DBR mirror consists of quater-wavelength thick high and low refractive index layers. The reflectance of DBR is proportional to the number of pairs and the difference between high and low index pairs \citep{m2}. The first and the last layers are AlGaAs. This enhances the coupling of light in/out of the structure since the refractive index of AlGaAs lies between those of GaAs and air \citep{m2}. 

\section{Conclusions}
In conclusion, we have analyzed the photon statistics of the light emitted by two optically coupled semiconductor micro-cavities each containing a quantum well. In view of conditional quantum dynamics, we have demonstrated that the field emitted by one QW can be controlled by the properties of the second QW. The steady state behaviour demonstrates that this proposed system can perform an efficient all optical switching. The optical switch can be made sensitive by tuning the properties of the two micro-cavities as well as the two quantum wells. The photon statistics is found to be controlled and tuned by appropriately changing the QW-cavity coupling and the photon tunneling rate. Furthermore, we have shown that the spectrum of the transmitted field consists of four distinct peaks in the strong coupling regime that corresponds to polariton resonances. In the intermediate, weak and extremely weak coupling regimes hybrid resonances appear. Our results demonstrate that the present scheme can, in principle, be used as a sensitive optical switch/optical sensors with the QW-cavity coupling and the tunneling rate as possible control parameters.

\begin{acknowledgements}
A. B. B acknowledges Birla Institute of Technology, Pilani for the financial support and facilities to carry out this research. S.A is grateful to Jawaharlal Nehru University, New Delhi for the Ph.D fellowship.
\end{acknowledgements}

\section{Appendix A}

\noindent Here we list the expressions, which appear in the equation for the intensity spectrum $S_{R}(\omega)$ (23):

\begin{equation}
f_{1}(\omega)= 2 \gamma J^{2} G_{1}^{2} [\gamma^2+(\Delta \omega_{2}-\omega)^2] n_{c L},
\end{equation}

\begin{equation}
f_{2}(\omega)= 2 \kappa J^{2} (\psi_{r}^{2}(\omega)+\psi_{i}^{2}(\omega)) n_{a L},
\end{equation}

\begin{equation}
f_{3}(\omega)= 2 \kappa G_{2}^{2} (\phi_{1r}^{2}(\omega)+\phi_{1i}^{2}(\omega)) n_{c R},
\end{equation}

\begin{equation}
f_{4}(\omega)=  \kappa \gamma (\phi_{1r}^{2}(\omega)+\phi_{1i}^{2}(\omega)) [\gamma^2+(\Delta \omega_{2}-\omega)^2] n_{a R},
\end{equation}

\begin{equation}
\psi_{r}(\omega)= \gamma^{2}-(\Delta \omega_{1}-\omega)(\Delta \omega_{2}-\omega),
\end{equation}

\begin{equation}
\psi_{i}(\omega)= \gamma (\Delta \omega_{1}+ \Delta \omega_{2}-2 \omega),
\end{equation}

\begin{equation}
\phi_{1r}(\omega)= G_{1}^{2}+ \kappa \gamma- (\Delta_{L}-\omega) (\Delta \omega_{1}-\omega),
\end{equation}

\begin{equation}
\phi_{1i}(\omega)=(\Delta_{L}-\omega) \gamma+(\Delta \omega_{1}-\omega) \kappa,
\end{equation}

\begin{equation}
\phi_{2r}(\omega) = G_{2}^{2}+\kappa \gamma - (\Delta_{R}-\omega) (\Delta \omega_{2}-\omega),
\end{equation}

\begin{equation}
\phi_{2i} (\omega) = (\Delta_{R}-\omega) \gamma +(\Delta \omega_{2}-\omega) \kappa.
\end{equation}


\begin{thebibliography}{99}

\bibitem{shields}
A. J. Shields, Nature Photonics \textbf{1}, 215 (2007).
%
\bibitem{ghosh}
S. Ghosh, W. H. W1ng, F. M. Mendoza, R. C. Myers, X. Li, N. Samarth, A. C. Gossard  and D. D. Awschaom,  Nature Materials, \textbf{5}, 261 (2006).
%
\bibitem{khitrova}
G. Khitrova, H. M. Gibbs, M. Khira, S. W. Koch, A. Scherer, Nature Physics, \textbf{2}, 81 (2006).
%
\bibitem{haicher}
X. Haicher, L. Furfaro, J. Javaoyes, M. Giudici, S. Balle, J. Tredicce, G. Tissoni, L. A. Lugiato, M. Brambilla and T. Maggipinto, Phys. Rev. A, \textbf{72}, 013815 (2005).
%
\bibitem{richard}
M. Richard, R. Romestian, R. Andre, L. S. Dang, Appl. Phys. Lett. \textbf{86}, 071916 (2005).
%
\bibitem{vahala}
K. Vahala, Optical Microcavities, Advanced Series in Applied Physics (World Scientific Publishing Company, 2005).
%
\bibitem{deveaud}
B. Deveaud, The Physics of Semiconductor Microcavities,  (WILEY VCH, 2007).
%
\bibitem{savona}
V. Savona, Physics of Semiconductor Microcavities: From Fundamentals to Nano Devices, edited by B. Deveaud, (WILEY VCH Verlag, 2007), Chap.1.
%
\bibitem{gibbs}
H. M. Gibbs, G. Khitrova and S. W. Koch, Nature Photonics, \textbf{5}, 273 (2011).
%
\bibitem{sete1}
E. A. Sete and H. Eluech, Phys. Rev. A, \textbf{82}, 043810 (2010).
%
\bibitem{sete2}
E. A. Sete, S. Das and  H. Eluech,  Phys. Rev. A, \textbf{83}, 023822 (2011).
%
\bibitem{sete3}
E. A. Sete, H. Eluech and S. Das, Phys. Rev. A, \textbf{84}, 053817 (2011).
%
\bibitem{sete4}
E. A. Sete and H. Eleuch, Phys. Rev. A, \textbf{85}, 043824 (2012).
%
\bibitem{sete5}
E. A. Sete, H. Eleuch and C. H. Raymond Ooi, Phys. Rev. A, \textbf{92}, 033843 (2015).
%
\bibitem{tassone}
F. Tassone, Y. Yamamoto, Phys. Rev. A, \textbf{62}, 063806 (2000).
%
\bibitem{yamamoto}
Y. Yamamoto et al., Semiconductor Cavity Quantum Electrodynamics (Springer, Berlin, 2000).
%
\bibitem{weisbuch}
C. Weisbuch, M. Nishioka, A. Ishikawa and Y. Arakawa, Phys. Rev. Lett., \textbf{69}, 3314 (1992).
%
\bibitem{pau}
S. Pau, G, Bjork, J, Jacobson, H, Cao and Y, Yamamoto, Phys. Rev. B, \textbf{51}, 14437 (1995).
%
\bibitem{jacobson}
J. Jacobson , S. Pau, H. Cao, G. Bjork and Y. Yamamoto Phys. Rev. A, \textbf{51}, 2542 (1995).
%
\bibitem{reithmaier}
J. P. Reithmaier, Semicond. Sci. Technol., \textbf{23}, 123001 (2008).
%
\bibitem{ishida}
N. Ishida, T. Byrnes, F. Nori and Y. Yamamoto, Sci. Rep., \textbf{3}, 1180: DOI:10.1038/srep 01180 (2015).
%
\bibitem{santhosh}
K. Santhosh, O. Bitton, L. Chuhtonov and G. Haran, Nat. Commun, \textbf{7}: 11823 DOI: 10.1038/ncomms 11823 (2016).
%
\bibitem{hennessy}
K. Hennessy, A. Badolato, M. Winger, D. Gerace, M. Atature, S. Gulde, S. Falt, E. L. Hu and  A. Imamoglu, Nature, \textbf{445}, 896 (2007).
%
\bibitem{chen}
Y. Chen, A. Traducci, F. Bassani, Phys. Rev. B, \textbf{52}, 1800 (1995).
%
\bibitem{sermage}
B. Sermage, S. Long, I. Abram, J. Y. Marzin, J. Bloch, R. Planel and V. Thierry-Mieg, Phys. Rev. B, \textbf{53}, 16516 (1996).
%
\bibitem{majumdar}
A. Majumdar, M. Bajcsy, D. Englund and J. Vu $\check{c}$ kovi$\acute{c}$, IEEE J. Selected Topics in Quantum Electronics, \textbf{18}, 1812 (2012).
%
\bibitem{kiraz}
A. Kiraz et al., J. Opt. B: Quantum Semiclass.Opt., \textbf{5}, 129 (2003).
%
\bibitem{faraon}
A. Faraon et al., New J. Phys., \textbf{13}, 055025 (2011).
%
\bibitem{lodahl}
P. Lodahl, Quantum Sci. Technol., \textbf{3}, 013001 (2018).
%
\bibitem{bhattacharya}
P. Bhattacharya, Proceedings of the IEEE, \textbf{95}, 1723 (2007).
%
\bibitem{bhattacherjee}
A. Bhattacherjee and M. Hasan, Journal of Modern Optics, DOI: 10.1080/09500340.2018.1455917, (2018).
%
\bibitem{heshami}
K. Heshamie, D. G. England,  P. C. Humphreys,  P. J. Bustard,  V. M. Acosta, J. Nunn and B. J. Sussman, J. Mod. Opt., \textbf{63}, 2005 (2016).
%
\bibitem{dory}
C. Dory, K. A. Fischer, K. Muller, K. G. Lagoudakis, T. Sarmiento, A. Rundquist, J. L. Zhang, Y. Kelaita and J. Vuckovic, Sci. Rep. \textbf{6}, 25172; DOI;10.1038/srep 25172 (2016).
%
\bibitem{englund}
D. Englund, A. Faraon, I. Fushman, N. Stoltz, P. Petroff and J. Vuckovic, Nature, \textbf{450}, 857 (2007).
%
\bibitem{robledo}
L. Robledo et al., Science, \textbf{320}, 772 (2008).
%
\bibitem{walls}
D. F Walls and G. J. Milburn, Quantum Optics, Springer-Verlag, Berlin, (2008).
%
\bibitem{qutip}
QuTip: Quantum Toolbox in Python, Version 4.2.0 (2017).
%
\bibitem{m1}
Jan Gudat, "Cavity Quantum Electrodynamics with quantum dots in microcavities", Phd. Thesis, University of Leiden (2012).
%
\bibitem{m2}
H. K. H Choy, "Design and fabrication of distributed Bragg reflectors for vertical-cavity surface-emitting lasers", M.Sc. Thesis, Mc Master University (1996).















































































\end{thebibliography}
\end{document}